\documentclass[11pt,aps,prd,preprint,preprintnumbers,showpacs,superscriptaddress,floatfix]{revtex4-1}
\usepackage{graphicx,}
\usepackage{graphicx,dcolumn,bm,epsfig,amsmath,amssymb,}
\usepackage{epsfig}
\usepackage{subfigure}
\usepackage{booktabs}
\usepackage{multirow}
\usepackage{float}
\usepackage{dcolumn}
\usepackage{bm}
\usepackage{amsmath}
\usepackage{amssymb}
\usepackage{bbm}
\usepackage{multirow}
\usepackage{setspace}
\usepackage{xcolor}
\usepackage[colorlinks,
            linkbordercolor={0 0 0},
            pdfborder={0 0 0},
            linkcolor=blue,
            citecolor=blue,
            urlcolor=blue,
            breaklinks=true]{hyperref}

\usepackage{tikz}
\usetikzlibrary{positioning,decorations.pathmorphing,decorations.markings,arrows}
\tikzset{
    vector/.style={decorate, decoration={snake}, draw},
    provector/.style={decorate, decoration={snake,amplitude=2.5pt}, draw},
    antivector/.style={decorate, decoration={snake,amplitude=-2.5pt}, draw},
    fermion/.style={draw=black,
      postaction={decorate},decoration={markings,mark=at position .55
        with {\arrow[draw=black]{>}}}},
    fermionbar/.style={draw=black, postaction={decorate},
                       decoration={markings,mark=at position .55 with {\arrow[draw=black]{<}}}},
    fermionnoarrow/.style={draw=black},
    gluon/.style={decorate, draw=black,decoration={coil,amplitude=4pt, segment length=6pt}},
    scalar/.style={dashed,draw=black,
      postaction={decorate},decoration={markings,mark=at position .55
        with {\arrow[draw=black]{>}}}},
    scalarbar/.style={dashed,draw=black,
      postaction={decorate},decoration={markings,mark=at position .55
        with {\arrow[draw=black]{<}}}},
    scalarnoarrow/.style={dashed,draw=black},
    electron/.style={draw=black,
      postaction={decorate},decoration={markings,mark=at position .55
        with {\arrow[draw=black]{>}}}},
    bigvector/.style={decorate, decoration={snake,amplitude=4pt}, draw},
}

\newlength{\x}
\newlength{\y}
\newlength{\z}
\phantomsection

\begin{document}
\preprint{IMSc/2021/02/01}

\title {Confinement-Deconfinement transition and $Z_2$ symmetry in
$Z_2+$Higgs theory}

\author{Minati Biswal}
\email{biswalmnt@gmail.com}              
\affiliation{Indian Institute of Science Education and Research, Mohali 
140306, India}

\author{Sanatan Digal}
\email{digal@imsc.res.in}              
\affiliation{The Institute of Mathematical Sciences, Chennai 600113, India}
\affiliation{Homi Bhabha National Institute, Training School Complex,
Anushakti Nagar, Mumbai 400085, India}

\author{Vinod Mamale}
\email{mvinod@imsc.res.in}
\affiliation{The Institute of Mathematical Sciences, Chennai 600113, India}
\affiliation{Homi Bhabha National Institute, Training School Complex,
Anushakti Nagar, Mumbai 400085, India}

\author{Sabiar Shaikh}
\email{sabiarshaikh@imsc.res.in}
\affiliation{The Institute of Mathematical Sciences, Chennai 600113, India}
\affiliation{Homi Bhabha National Institute, Training School Complex,
Anushakti Nagar, Mumbai 400085, India}

\begin{abstract}
We study the Polyakov loop and the $Z_2$ symmetry in the lattice $Z_2+$Higgs theory in four dimensional space using 
Monte Carlo simulations. The results show that  this symmetry is realised in the Higgs symmetric phase for large 
number of ``temporal'' lattice sites. To understand this dependence on the number of ``temporal'' sites, we consider a 
one dimensional model by keeping terms of the original action corresponding to a single spatial site. In this 
approximation the partition function can be calculated exactly as a function of the Polyakov loop. The 
resulting free energy is found to have the $Z_2$ symmetry in the limit of large temporal sites. We argue
that this is due to $Z_2$ invariance as well as dominance of the distribution or density of states corresponding to the action.
\end{abstract}

\maketitle

\section{Introduction}


$Z_N$ symmetry plays an important role in the confinement-deconfinement (CD) transition in pure $SU(N)$ gauge 
theories~\cite{tHooft:1977nqb, McLerran:1981pb, Belyaev:1991gh}. In these theories, at finite temperature, the allowed 
gauge transformations are classified by the centre of the gauge group, i.e $Z_N$. Under these $Z_N$ gauge 
transformations, i.e $Z_N$ symmetry, the Polyakov loop ($L$) transforms like magnetisation, in spin models~\cite{Kogut:1979wt}. 
In the confinement and deconfinement phases the Polyakov loop acquires vanishing and non-zero thermal average values 
respectively, hence plays the role of an order parameter for the confinement-deconfinement (CD) transition,~\cite{Kuti:1980gh, 
McLerran:1980pk, Weiss:1980rj, Creutz:1980zw,Svetitsky:1982gs}. In the deconfined phase, the $Z_N$ symmetry 
is spontaneously broken which leads to $N$-degenerate states~\cite{Yaffe:1982qf, Svetitsky:1985ye, Celik:1983wz}.


The $Z_N$ symmetry of pure $SU(N)$ gauge theory is spoiled when matter fields are included. Gauge transformations
which are not periodic in temporal directions can not act on the matter fields. These may act only on the gauge fields but in
the process the action does not remain invariant. There are many
studies on the effect of matter fields on this symmetry. Perturbative loop calculations of the Polyakov loop
effective potential show that this symmetry is explicitly broken by matter fields in the fundamental representation
~\cite{Weiss:1981ev, Belyaev:1991np, Guo:2018scp}. The
mean-field approximations  of lattice partition functions in the strong coupling limit also show the explicit breaking
of the $Z_N$ symmetry~\cite{Green:1983sd, Karsch:2000zv}. On the other hand, non-perturbative studies of $CD$ transition in
$2-$colour QCD show a sharp transition suggesting small explicit breaking of $Z_2$ symmetry~\cite{Satz:1985js}. 

Recent non-perturbative Monte Carlo simulations of $SU(2)+$Higgs theory show that the strength of $Z_2$ explicit 
breaking depends on the Higgs condensate~\cite{Biswal:2015rul}. These studies find that the $CD$ transition 
exhibits critical behaviour in the Higgs symmetric phase for large number of temporal sites 
($N_\tau$)~\cite{Biswal:2015rul}. 
The distributions of the Polyakov loop are found to be $Z_2$ symmetric, albeit within statistical 
errors, suggesting the realisation of $Z_2$ symmetry in the Higgs symmetric phase. In reference~\cite{Biswal:2016xyq}, 
it was argued that the emergence of $Z_2$ symmetry is due to enhancement of the configuration/ensemble space 
with $N_\tau$. This enhancement makes it possible that the change in the Euclidean action due to $Z_2$ ``rotation" 
of gauge links can be compensated by changing the Higgs field appropriately. This was numerically tested by updating
the Higgs field using Monte Carlo steps after $Z_2$ rotating the gauge fields.\\

The non-invariance of the action under $Z_2$ gauge transformation which are not periodic in temporal directions
does not necessarily imply the explicit breaking of $Z_2$ symmetry. The presence of $Z_2$ symmetry or it's explicit
breaking can only be inferred from the free energy of the Polyakov loop. In the free energy or the partition 
function calculations, two factors play important roles. They are the distribution of the action, which is also known as the density
of states ($DoS$) and the Boltzmann factor. The latter clearly does not respect the $Z_2$ symmetry. So the realisation of 
the $Z_2$ symmetry must come from the 
$DoS$ and it's dominance over the Boltzmann factor. Computing the $DoS$ in $SU(N)+$Higgs theory is a 
difficult task as the configuration space is infinite. In this situation, the $Z_2+$Higgs theory in four dimensions provides a 
suitable alternative. Since the field variables take values $\pm1$, it is possible to calculate the $DoS$ with some simplifications. 

The $Z_2+$Higgs theory has been extensively studied in literature~\cite{Balian:1974ts,Balian:1974xw,Creutz:1979he, Fradkin:1978dv, Callaway:1981rt, Bhanot:1981ug, Bhanot:1980ga}. The phase diagram of this theory is found to be similar to that of $SU(N)+$Higgs theories in $3$ and $4-$dimensions ~\cite{Jongeward:1980wx,Creutz:1983ev}. Though, in this theory
there is no analog of the beta-functions of $SU(N)+$Higgs theories and the temperature is controlled by the couplings of the
therory \cite{Caselle:1995wn}. The similarity with the $SU(N)+$Higgs theories arises when
periodic/anti-periodic boundary condition is imposed on the Higgs field, in any one of the four dimensions. As a consequence
gauge transformations which are not periodic in this ``temporal'' direction are not allowed and the $Z_2$ symmetry is explicitly broken similar to the explicit breaking of $Z_N$ symmetry in $SU(N)+$Higgs theories. It is important to note the difference 
on the role of $N_\tau$ between $Z_2+$Higgs and $SU(N)+$Higgs theories. Though in both cases increase in $N_\tau$ introduces additional degrees of freedom, in $SU(N)+$Higgs theory to study $Z_N$ symmetry at fixed temperature the couplings need to be tuned. 

In this paper, the $Z_2$ symmetry of the Polyakov loop and the nature of $CD$ transition are studied by varying the number 
of lattice points, $N_\tau$, along the temporal direction. The computations are mostly done on the Higgs symmetric 
side of the Higgs transition line. Our results show that the $Z_2$ symmetry is realised for large $N_\tau$. Also
the behaviour of the $CD$ transition is found to be similar to the pure gauge case apart from the location of the critical
point. To understand the role of $N_\tau$ a $0+1$ dimensional model is considered by keeping temporal component of the 
gauge Higgs interaction corresponding to a single spatial coordinate. The reason for this choice is the fact that only
the  temporal component of the gauge Higgs interaction is sensitive to the $Z_2$ gauge transformations.
For the one dimensional model the Polyakov loop
can take values $\pm 1$. For each of these cases the free energy can be calculated exactly. The free energy calculations
show the emergence of $Z_2$ symmetry in the large $N_\tau$ limit for arbitrary interaction coupling. Further 
the Monte Carlo results for the distribution of the interaction term is reproduced well by $0+1$ dimensional $DoS$ 
with a simple Boltzmann factor, though with a different value of the coupling strength. The $DoS$ for both
values of the Polyakov loop is sharply peaked at zero. $Z_2$ symmetry is clearly observed near the peak, the 
differences appear when the action takes the limiting values. Since the peak height grows with $N_\tau$, the $DoS$
will dominate the thermodynamics in the $N_\tau\to \infty$, leading to vanishingly small $Z_2$ explicit symmetry breaking.

This paper is organised as follows. In section II, we discuss the $Z_2$ symmetry in $Z_2+$Higgs theory. This is followed by 
numerical simulations of $CD$ transition and the $Z_2$ symmetry in pure gauge theory and in the presence of Higgs in section III. 
In section IV, we derive the free energy of the Polyakov loop in a $0+1$ dimensional model, and relate
the results to  $4-$dimensional Monte Carlo simulations. In section V, discussions and conclusions are presented.
\section{ $Z_2$ symmetry in $Z_2+$Higgs gauge theory. }
\begin{figure}[H]
 \centering
  \includegraphics[width=0.3\textwidth]{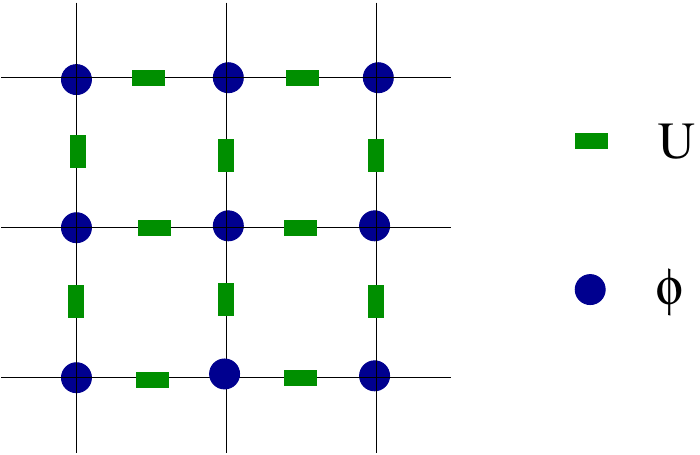}
 \caption{Position of gauge links U and Higgs fields $\Phi$ on lattice}
\label{fig1}
\end{figure}

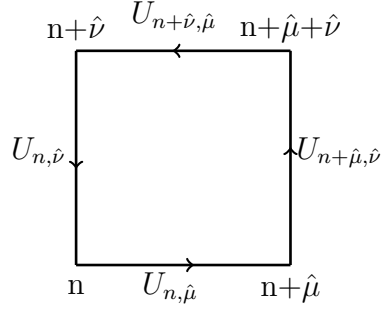
\begin{figure}[h!]
\begin{center}
\begin{tikzpicture}[line width=1 pt, scale=0.5]
\draw[fermion] (0,0) -- (5.7,0);
\draw[fermion] (5.7,0) -- (5.7,5.7);
\draw[fermion] (5.7,5.7) -- (0,5.7);
\draw[fermion] (0,5.7) -- (0,0);
\node at (0,-0.6) {n};
\node at (5.7,-0.6) {n+$\hat{\mu}$};
\node at (5.7,6.3) {n+$\hat{\mu}$+$\hat{\nu}$};
\node at (0,6.3) {n+$\hat{\nu}$};
\node at (2.5,-0.6) {$U_{n,\hat{\mu}}$};
\node at (7.0,3.0) {$U_{n+\hat{\mu},\hat{\nu}}$};
\node at (2.6,6.6) {$U_{n+\hat{\nu},\hat{\mu}}$};
\node at (-1.0,3.0) {$U_{n,\hat{\nu}}$};
\end{tikzpicture}
\end{center}
\caption{Sketch of an elementary plaquette $U_P$}
\label{fig2}
\end{figure}

\noindent 
The action for the $Z_2+$Higgs theory in four dimensional lattice $(N_s^3 \times N_\tau)$ is given by,
\begin{equation}
S = -\beta_g\sum_P  U_P - \kappa\sum_{{n},\hat{\mu}} 
\Phi_{{ n}+\hat{\mu}}U_{n,\hat{\mu}}\Phi_{n}.
\label{action}
\end{equation}
Here $n=(n_1,n_2,n_3,n_4)$ represents a point on the lattice with $1\le n_1,n_2,n_3\le N_s$ and $1\le n_4\le N_\tau$. As mentioned
above we assume that the fourth direction is the temporal direction.
$U_{n,\hat{\mu}}$ represents the gauge links in $\hat{\mu}$ direction between the lattice point $n$ and $n+\hat{\mu}$. The Higgs 
field $\Phi_{n}$ lives at the site $n$. Both $U_{n,\hat{\mu}}$ and $\Phi_n$ take values $\pm 1$. $\beta_g$ is the gauge coupling 
and $\kappa$ is the gauge Higgs interaction strength. Figure.~\ref{fig1} shows a schematic layout of the gauge links and 
Higgs variable on the lattice.
\noindent
The plaquette $U_P$ which is path ordered product of the links along an elementary square on the $\mu-\nu$ plane, i.e
\begin{equation}
U_P=U_{n,\hat{\mu}} U_{n+\hat{\mu},\hat{\nu}} U_{n+\hat{\nu},\hat{\mu}} U_{n,\hat{\nu}}.
\end{equation}
Figure.~\ref{fig2} shows the sketch of an elementary plaquette.
\noindent
The pure gauge part of the action, first term in Eq.\ref{action}, is invariant under the $Z_2$ gauge transformations,
\begin{equation}
U_{n,\hat{\mu}} \rightarrow V_{n} U_{n,\hat{\mu}} V^{-1}_{n+\hat{\mu}}
\end{equation}
where $V_n=\pm 1\in Z_2$. The $V_n$'s satisfy the following boundary 
condition,
\begin{equation}
V(\vec{n},n_4=1) = zV(\vec{n},n_4=N_\tau).
\end{equation}
$z=\pm 1\in Z_2$. So the gauge transformations can be classified by the group $Z_2$. For $z=-1$ the
gauge transformations are anti-periodic in the temporal direction.

The Polyakov loop, which is defined as the product of links along the temporal direction, i.e,
\begin{equation}
L(\vec{n}) = \displaystyle\prod_{n_4=1}^{N_\tau} U_{{(\vec{n},n_4)},\hat{4}}
\label{plp}
\end{equation}
transforms non-trivially under $Z_2$ gauge transformations~\cite{Svetitsky:1982gs}. 
It is easy to see that the Polyakov loop transforms as,
\begin{equation}
L(\vec{n}) \rightarrow zL(\vec{n}).
\end{equation}
This transformation property of the Polyakov loop under $Z_2$ (or $Z_N$ in general) gauge 
transformation is similar to that of magnetisation in the Ising model. The partition function in the pure gauge case ($\kappa=0$) 
is given by,
\begin{equation}
{\cal Z}=\int DU e^{-S}.
\end{equation}
Since the action for $\kappa=0$ is invariant under $Z_2$ gauge transformations, any configuration and it's gauge rotated 
counterpart will contribute equally to the partition function. Therefore the distribution of the Polyakov loop exhibits $Z_2$ 
symmetry in this case. Equivalently the free energy of the Polyakov loop will have $Z_2$ symmetry.\\

The presence of the Higgs field changes the space of allowed gauge transformations. The reason being that the Higgs field is 
required to be periodic in the temporal direction. Under a gauge transformation, $\Phi_n$ transforms as,
\begin{equation}
\Phi_n \rightarrow V_n\Phi_n.
\end{equation}
Now the periodic boundary condition of $\Phi$ would be spoiled if non-periodic gauge transformations, characterised by 
$z=-1$ are allowed. In this case given a configuration, one can define a $Z_2$ counterpart in which only the gauge links 
are $Z_2$ rotated. Obviously these pair of configurations will not contribute equally to the partition function for 
$\kappa\ne 0$. So according to the Boltzmann factor, $\sum_{\vec{n}} L(\vec{n})$ and $-\sum_{\vec{n}}L(\vec{n})$ are
non degenerate. This situation is similar to the presence of an external field in the Ising model. However, the status of $Z_2$ 
symmetry in the free energy can be answered only after integrating out the Higgs field for a given $L(\vec{n})$ and it's
$Z_2$ rotated configurations.

The Polyakov loop and Ising spins are similar in how they transform under respective transformations. However there is an 
important difference between them. This becomes clear when one compares $L(\vec{n})$ and an Ising spin at a spatial point 
$\vec{n}=\{n_1,n_2,n_3\}$. A given value of $L(\vec{n})$  is associated with an entropy factor. This is because there are many 
different combinations of $U_{(\vec{n},n_4),\hat{4}}$ and $\Phi_{\vec{n},n_4}$ are possible for a given value of $L(\vec{n})$. 
Larger the $N_\tau$, larger is the corresponding entropy. This aspect of the Polyakov loop needs to be taken into account
to understand the explicit breaking or realisation of $Z_2$ ($Z_N$) symmetry, which is done in section IV. In the following 
section, we discuss the algorithm of the Monte Carlo 
simulations~\cite{Creutz:1979kf}, present simulation results for the phase diagram in the $\beta_g -\kappa$ plane, distribution of 
the Polyakov loop and $CD$ transition in the Higgs symmetric phase etc.

\section{Numerical technique and Monte Carlo simulation results.}

In the Monte Carlo simulations, the Metropolis algorithm is used for sampling the statistically significant 
configurations~\cite{Hastings:1970aa}. To update a particular gauge link $U_{n,\mu}$, we consider the change in 
the action by flipping it. If the action decreases then the flipped gauge link is accepted for the new configuration. If the
action increases by $\Delta S$ then the new link is accepted with probability $Exp(-\Delta S)$. 
The same procedure is adopted for $\Phi_n$. The process of updating 
is carried out over all $n$ and $\mu$ in multiple sweeps. Configurations separated by $10$ sweeps are used in our analysis, 
which brings 
down the autocorrelation between successive configurations to an acceptable level. For this simulations,
$N_\tau=4-24$ and $N_s=16-84$ with $N_s/N_\tau = 4$ lattices have been considered~\cite{Engels:1981ab}.\\

\begin{figure}[h!]
\centering
  \subfigure
  {\rotatebox{360}{\includegraphics[width=0.48\hsize]
      {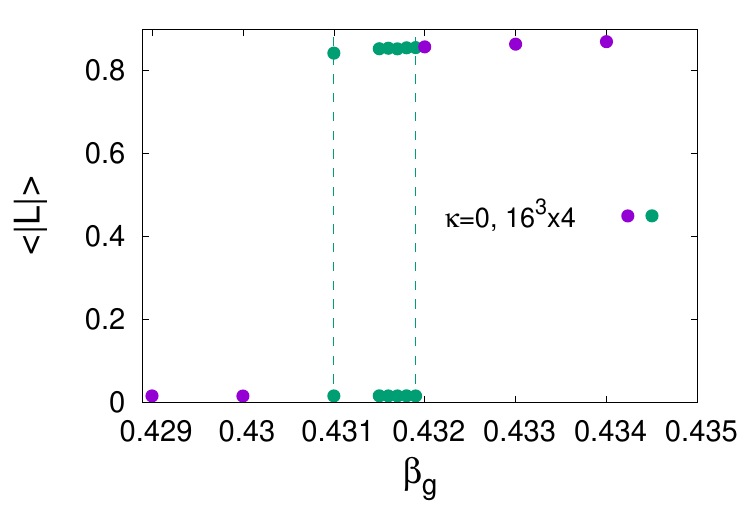}}
  }
  \subfigure
  {\rotatebox{360}{\includegraphics[width=0.48\hsize]
      {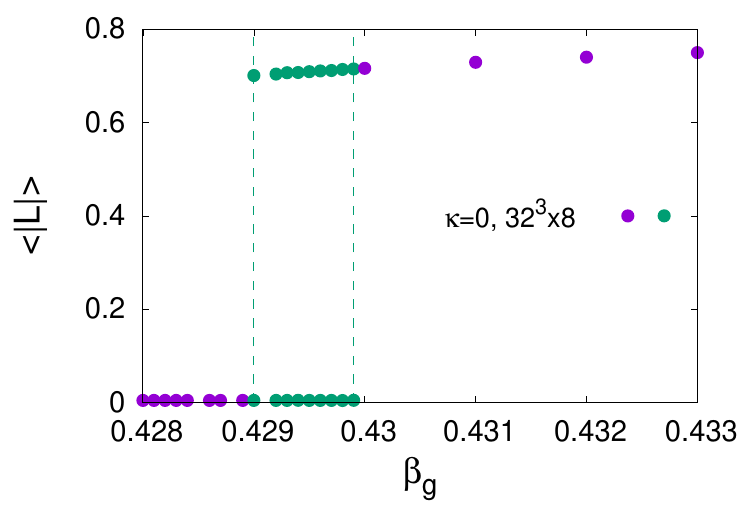}}
  }
\caption{The average of the Polyakov loop vs $\beta_g$ for $N_\tau =$ 4 and 8.}
\label{fig3}
\end{figure}

The pure gauge simulations are initially performed to understand the nature of $CD$ transition and $Z_2$ symmetry
of the Polyakov loop. The simulations were repeated in the presence of $\Phi$ to study its effects. The pure gauge transition
has been studied previously in the mean-field approximations \cite{Balian:1974ts}, which finds the transition is first order in four dimensions. Also using duality transformations it can be shown that the critical $\beta_g\sim0.4407$ for
$\kappa=0$ \cite{Balian:1974xw}. These results are supported by Monte Carlo simulations of smaller lattices \cite{Creutz:1979he}.
The simulations carried out in this work are also consistent with these results. In figure.~\ref{fig3} 
the average of the Polyakov loop is plotted $vs$ $\beta_g$ for $N_\tau=4,8$. There is a range in $\beta_g$ for which
clearly separated peaks in the distribution of the Polyakov loop has been observed. We take average of the Polyakov loop values
corresponding to each peak separately. Therefore we have two points in the figure for a given $\beta_g$. The two peaks
also suggest that the transition is first order. For larger lattice sizes the range of $\beta_g$ over 
which two states are observed increases~\cite{Damgaard:1986jg}. This is expected as strength of fluctuations relatively 
decrease with volume ( when 
correlation length is smaller than the spatial size of the system), making it difficult for the field climb over the barrier
and cross to the other side. 

\begin{figure}[h]
\centering
  \subfigure
  {\rotatebox{0}{\includegraphics[width=0.60\hsize]
      {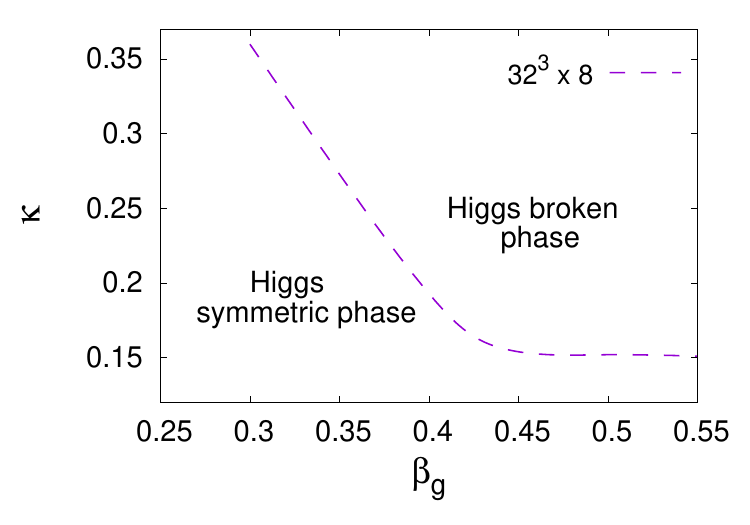}}
  }
\caption{Phase diagram}
\label{fig6}
\end{figure}

The effect of the $\Phi$ field on the $CD$ transition and $Z_2$ symmetry is expected to depend on $\kappa$. To relate these
two aspects of pure gauge theory to the phases of the Higgs field, simulations were performed to obtain the Higgs
transition line. For a given $\beta_g$, $\kappa>\kappa_c$ corresponds to
the Higgs broken phase. In this phase the action term dominates. For $\kappa < \kappa_c$ the fluctuations of the Higgs
rather than the action dominate the thermodynamic properties. This situation is similar to the Ising model at high 
temperatures.  In Fig.\ref{fig6} the Higgs transition line is plotted in the $\beta_g -\kappa$ plane. The location of the phase 
boundary is obtained by studying the $\kappa$ dependence of the interaction term and it's fluctuations
for different values of  $\beta_g$. In our simulations the Higgs transition is found to be first order for intermediate range of
$\beta$ and crossover for both small and large $\beta$, as observed in previous studies~\cite{Jongeward:1980wx, Creutz:1983ev}. 
For  large $\beta_g$ critical $\kappa_c$ remains flat and increases with $\beta_g$ in the small $\beta_g$ range. In our
simulations the critical values ($\beta_c,\kappa_c$) were found to vary mildly with $N_\tau$.

\begin{figure}[h]
\centering
  \subfigure
  {\rotatebox{360}{\includegraphics[width=0.48\hsize]
      {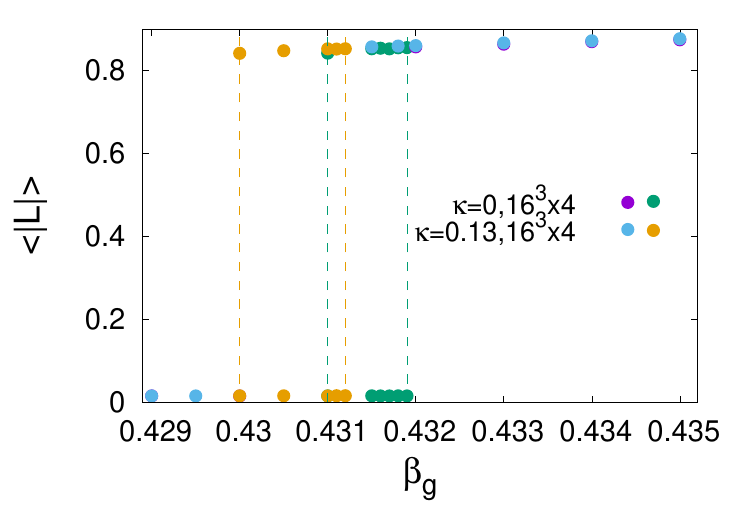}}
  }
  \subfigure
  {\rotatebox{360}{\includegraphics[width=0.48\hsize]
      {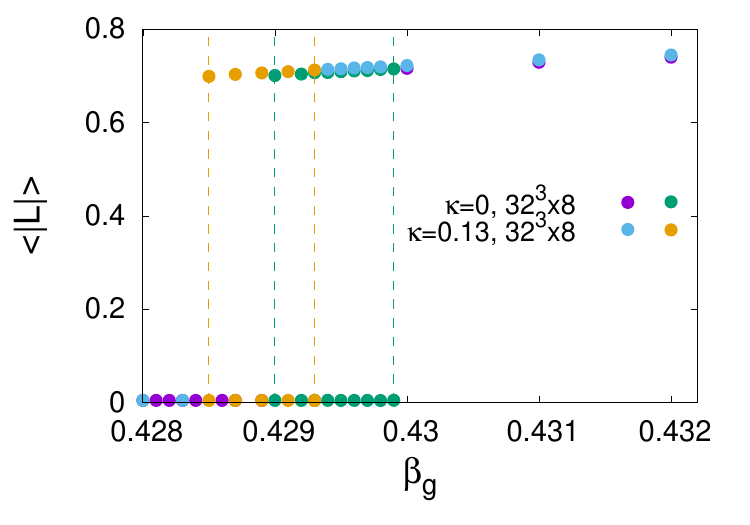}}
  }
\caption{The average of the Polyakov loop vs $\beta_g$ for $N_\tau =$ 4 and 8.}
\label{fig8a}
\end{figure}

In the Higgs broken phase, i.e large $\kappa$, the interaction term dominates over the entropy. The action takes the
largest value when all the temporal links are $+1$. So it is expected that in the Higgs phase $Z_2$ symmetry is
badly broken, also observed in our simulations. In the Higgs symmetric phase, it is the fluctuations of Higgs in other 
words the distribution of the interaction term dominate. In this phase there is a possibility for realisation of $Z_2$ symmetry. In 
figure.~\ref{fig8a} we show $CD$ transition in the Higgs symmetric phase $(\kappa=.13)$. For comparison, $\kappa=0$ results also have 
been included. The $CD$ transition is first order even in the presence of $\Phi$, though the transition point shifts
to lower values of $\beta_g$.

\begin{figure}[!tbp]
  \centering
  \begin{minipage}[b]{0.45\textwidth}
    \includegraphics[width=\textwidth]{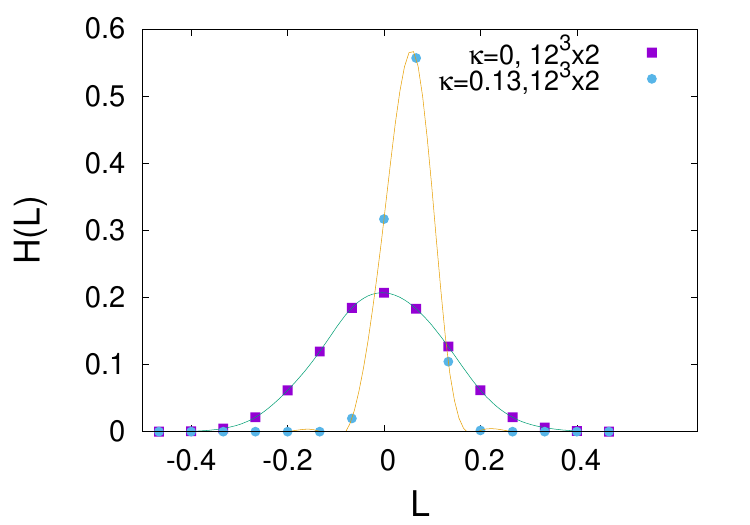}
    \caption{Histogram of L in the confined phase.}
 \label{fig8}
  \end{minipage}
   \hspace{0.3cm}
  \begin{minipage}[b]{0.45\textwidth}   
    \includegraphics[width=\textwidth]{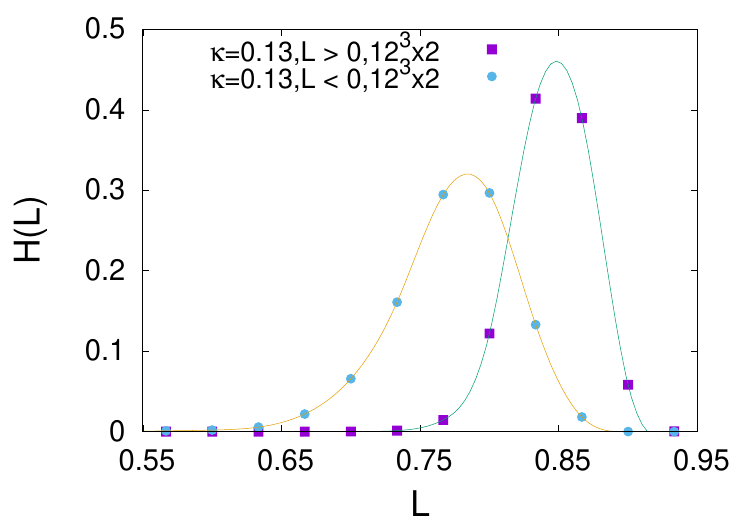}
    \caption{Histogram of L in the deconfined phase.}
   \label{fig9}
  \end{minipage}
\end{figure}

\begin{figure}[!tbp]
 \centering
  \begin{minipage}[b]{0.45\textwidth}
    \includegraphics[width=\textwidth]{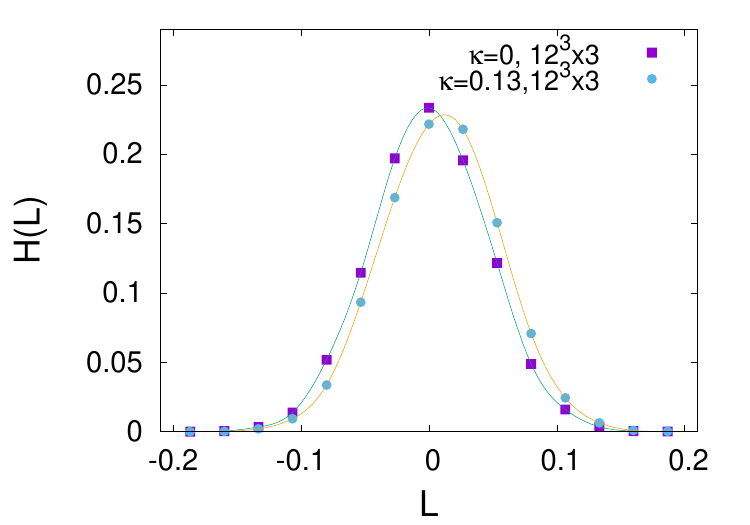}
    \caption{Histogram of L in the confined phase.}
    \label{fig10}
  \end{minipage}
   \hspace{0.3cm}
  \begin{minipage}[b]{0.45\textwidth}
    \includegraphics[width=\textwidth]{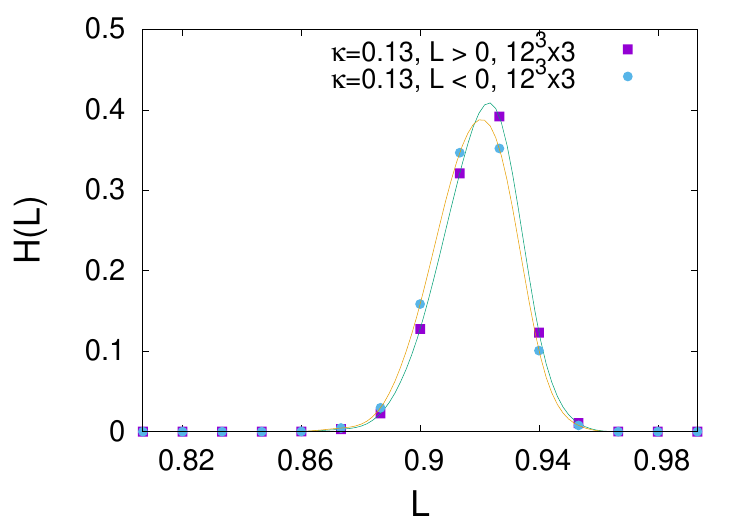}
    \caption{Histogram of L in the deconfined phase.}
   \label{fig11}
  \end{minipage}
  \end{figure}

\begin{figure}[!tbp]
\centering
 \begin{minipage}[b]{0.45\textwidth}
    \includegraphics[width=\textwidth]{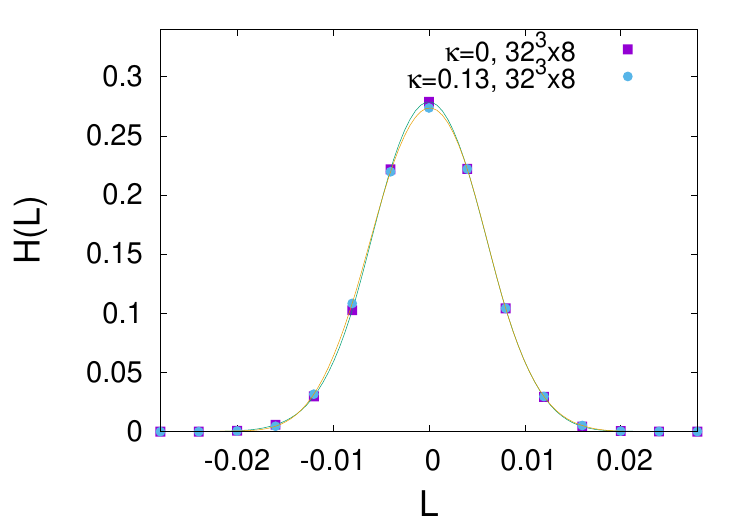}
    \caption{Histogram of L in the confined phase.}
   \label{fig51}
  \end{minipage}
   \hspace{0.3cm}
  \begin{minipage}[b]{0.45\textwidth}
    \includegraphics[width=\textwidth]{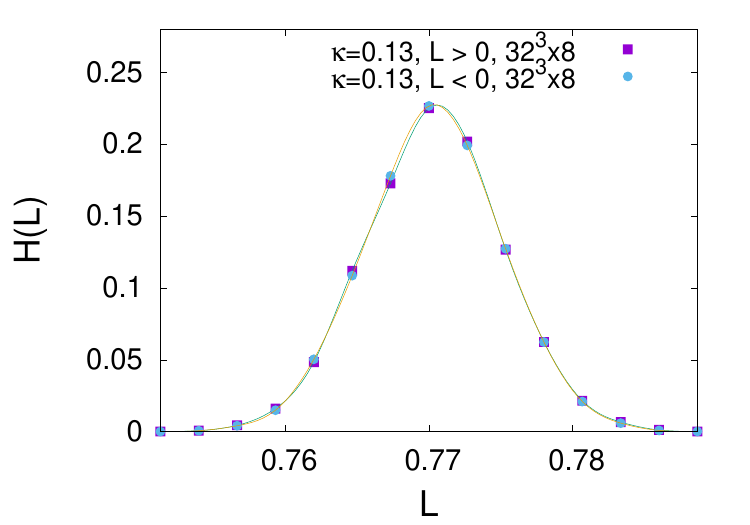}
    \caption{Histogram of L in the deconfined phase.}
   \label{fig61}
  \end{minipage}
 \end{figure}

To check the $N_\tau$ dependence of the $Z_2$ symmetry at $\kappa=.13$, the distribution of Polyakov loop is computed both
in the confined and the deconfined phases for $N_\tau=2,3,$ and 8. In the deconfined phase, $L < 0$ data is $Z_2$ rotated and then compared with $L > 0$ data. The distributions/histograms are shown in figures.~\ref{fig8}-\ref{fig61}. For
$N_\tau =2$ the histograms clearly show there is no $Z_2$ symmetry. In the deconfinement side there is no $Z_2$ symmetry as the two Polyakov loop sectors do not overlap. For $N_\tau=3$ the two peaks corresponding to the two sectors are approaching towards each other. For $N_\tau=8$, the histogram of Polyakov loop for two $Z_2$ sectors agree well with  each other.

\begin{figure}[!tbp]
  \centering
  \begin{minipage}[b]{0.45\textwidth}
    \includegraphics[width=\textwidth]{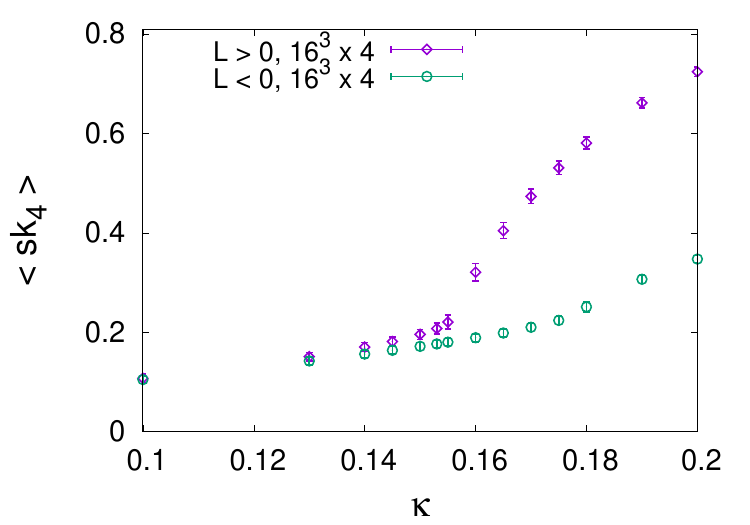}
    \caption{$sk_4$ average vs $\kappa$ for $\beta_g=0.435$ on $16^3\times  4$ lattice}
	  \label{fig12}
  \end{minipage}
\hspace{0.3cm}
  \begin{minipage}[b]{0.45\textwidth}
    \includegraphics[width=\textwidth]{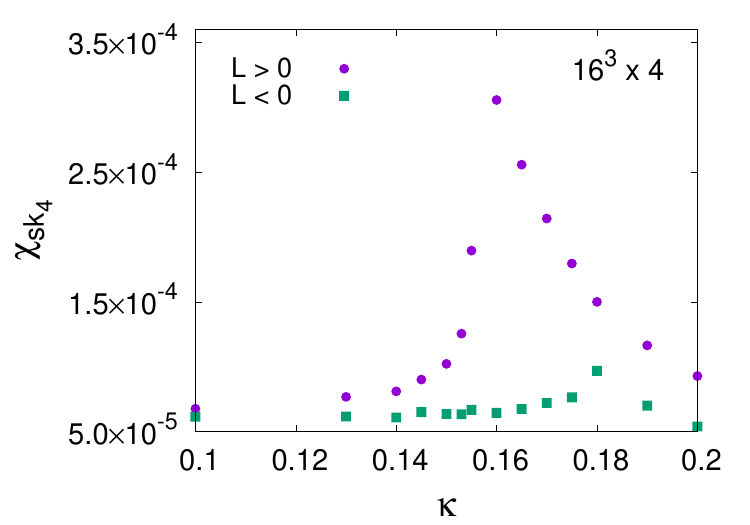}
    \caption{$sk_4$ fluctuation vs $\kappa$ for $\beta_g=0.435$ on $16^3\times  4$ lattice }
	  \label{fig13}
  \end{minipage}

  \centering
  \begin{minipage}[b]{0.45\textwidth}
    \includegraphics[width=\textwidth]{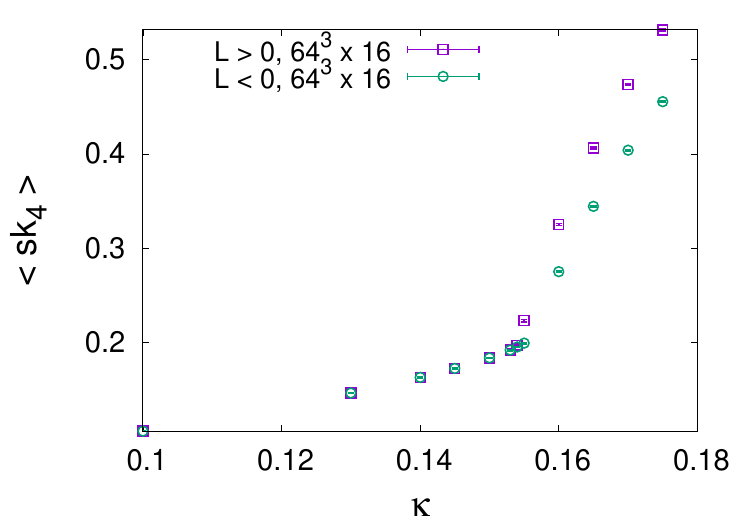}
    \caption{$sk_4$ average vs $\kappa$ for $\beta_g=0.435$ on $64^3\times  16$ lattice }
	  \label{fig14}
  \end{minipage}
   \hspace{0.3cm}
  \begin{minipage}[b]{0.45\textwidth}
    \includegraphics[width=\textwidth]{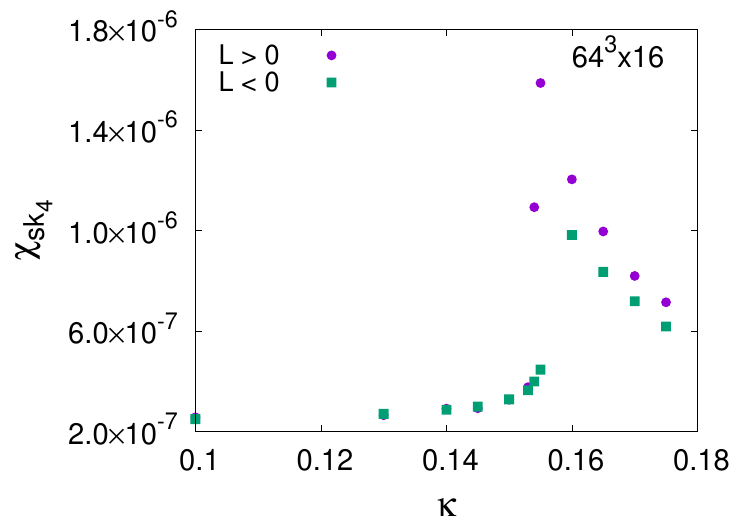}
    \caption{$sk_4$ fluctuation vs $\kappa$ for $\beta_g=0.435$ on $64^3\times  16$ lattice }
	  \label{fig15}
  \end{minipage}
 \end{figure}

The $\kappa$ dependence of the $Z_2$ symmetry is studied by computing the thermal average of the temporal part of the 
interaction, i.e $ sk_4=\sum_{n} \Phi_n U_{n,\hat{4}} \Phi_{n+\hat{4}}^\dag$ and the corresponding susceptibility $\chi_{sk_4}$.
These simulations are carried out in the deconfined phase, as there are two $Z_2$ states corresponding to each sector
of the Polyakov loop. The results for ($\left<sk_4\right>,\chi_{sk_4}$) are shown in figures.~\ref{fig12}-\ref{fig15}. For all 
$N_\tau$ values the difference in ($\left<sk_4\right>,\chi_{sk_4}$) for these two sectors is vanishingly small for small enough
$\kappa$. For larger $N_\tau$, the kappa value at which the two polyakov loop sectors differ significantly 
in $sk_4$ and $\chi_{sk_4}$ is higher. For the largest considered, $N_\tau=24$ , the two sectors agree in 
($\left<sk_4\right>,\chi_{sk_4}$) up to the Higgs crossover point $\kappa<\kappa_c$. When Higgs transition 
is first order the $Z_2$ symmetry is observed in the Higgs symmetric phase even for $\kappa>\kappa_c$. Note that
for $\kappa>\kappa_c$ the Higgs symmetric phase is meta-stable.\\

It is clear from our $3+1$ dimensional simulations that the $Z_2$ symmetry is realised in the Higgs symmetric phase 
for large $N_\tau$, i.e the partition function averages of physical observables exhibit the $Z_2$ symmetry. Though we have focussed mostly on simulations around the Higgs transition region, it is important
to look at the consequence of the $Z_2$ symmetry realisation across the CD transition. The CD transition line originates from the $\beta-$axis, runs parallel to $\kappa-$axis for small $\kappa$. For larger $\kappa$, $\beta_g=\beta_c$ decreases and the transition line merges with the Higgs transition line~\cite{Jongeward:1980wx, Creutz:1983ev}. For small but non-zero $\kappa$ the CD transition is first order for $N_\tau\ge 3$. For large $N_\tau$, for $\beta_g\ge\beta_c$ the Polyakov loop distribution shows two $Z_2$ symmetric peaks. In the confined phase,  for $\beta_g \le \beta_c$, the distribution exhibits a single peak. With increase in $N_\tau$ the peak of the distribution steadily approaches $L=0$ and
simultaneously the distribution exhibiting $Z_2$ symmetry.

For large $N_\tau$, below $\beta_c(N_\tau)$ thermal average of the Polyakov loop $\langle L \rangle = 0$. Note that $\langle L \rangle \propto Exp(-F/T)$, where $F$ is the free energy between static charges. This suggests that for $\beta \le \beta_c(N_\tau)$ static charges are confined. Previously confinement was observed only in the $\beta\to 0$ limit \cite{Fradkin:1978dv}.  It would be interesting to study the confinement aspects of the $Z_N$ symmetry realisation in $SU(N)$ gauge theories. For fixed $\kappa$ and $\beta$ in the confinement phase we observe that the free energy $F$ saturates for large $N_\tau$. This implies that the approach $\langle L \rangle \to 0$ is merely due to the temperature $T \to 0$.

 To understand the realisation of $Z_2$ symmetry in the current theory, we consider a $0+1$ dimensional model keeping
 only the temporal component of the interaction term corresponding to a single spatial coordinate in the following section.
 
\section{The partition function and density of states in $0+1$ dimensions}

The temporal component of the gauge Higgs interaction corresponding to a particular spatial site can be
written as,
\begin{equation}
S_{1D} =-\kappa sk_4,~~~sk_4= \sum_{n=1}^{N_\tau} \Phi_n U_n \Phi_{n+1}.
\end{equation}
 $n$ denotes the temporal lattice site, i.e $1 \le n\le N_\tau$. $\Phi_n$ satisfies the periodic boundary condition 
 $\Phi_{N_\tau+1}=\Phi_1$. Since the action will not be invariant if a $z=-1$ gauge transformation is made on $U_i$'s, the action 
 breaks the $Z_2$ symmetry explicitly. For this model the Polyakov loop can take values $\pm 1$. To see the $N_\tau$
 dependence of the $Z_2$ symmetry we calculate the free energy $V(L,N_\tau)$. To simplify the
 calculations we set $U_i=1$, for $i=1,2,...N_\tau-1$ and $U_{N_\tau} = L$. All other configurations of $U_i$ corresponding to a 
 given value of $L$ are gauge equivalent. Now the partition  function for $L=1$ is nothing but that of the one dimensional Ising chain. For $L=-1$ the only difference is that the coupling between $\Phi_{N_\tau}$ and $\Phi_1$ is anti-ferromagnetic. For each choice of $L$ the partition  function can be calculated exactly, i.e,
 \begin{equation}
 {\cal Z}(L=1) = \lambda_1^{N_\tau} + \lambda_2^{N_\tau},~~{\cal Z}(L=-1) = \lambda_1^{N_\tau} - \lambda_2^{N_\tau},
 \end{equation}
 where $\lambda_1 = e^\kappa+e^{-\kappa}$ and $\lambda_2 = e^\kappa-e^{-\kappa}$. The corresponding free energies in the large $N_\tau$ limit are given by, 
  \begin{equation}
	  V(L=1)=V(L=-1)=-TN_\tau {\rm log}(\lambda_1).
 \end{equation}
 This results show that there is $Z_2$ symmetry in $0+1$ dimensions in the limit of $N_\tau\to \infty$. 
 
   \begin{figure}[!tbp]
\centering
    \begin{minipage}[b]{0.45\textwidth}
    \includegraphics[width=\textwidth]{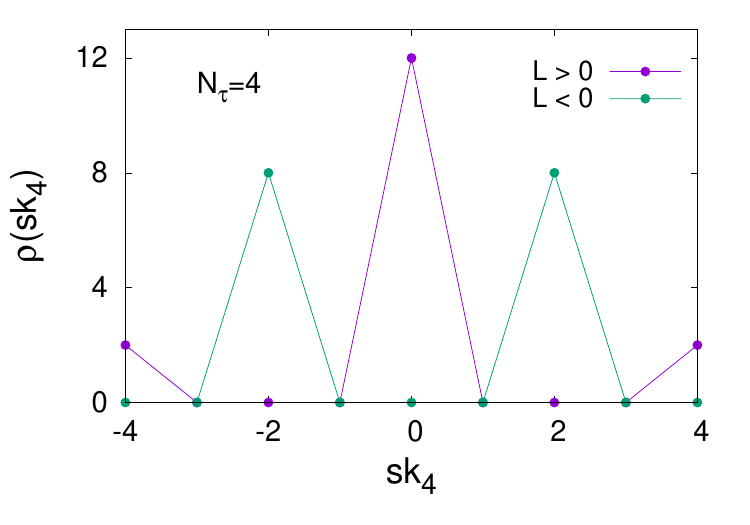}
    \caption{$\rho(sk_4)$ for $\kappa=0$ in 0+1}
   \label{fig22}	    
  \end{minipage}
  \hspace{0.3cm}
  \begin{minipage}[b]{0.45\textwidth}
    \includegraphics[width=\textwidth]{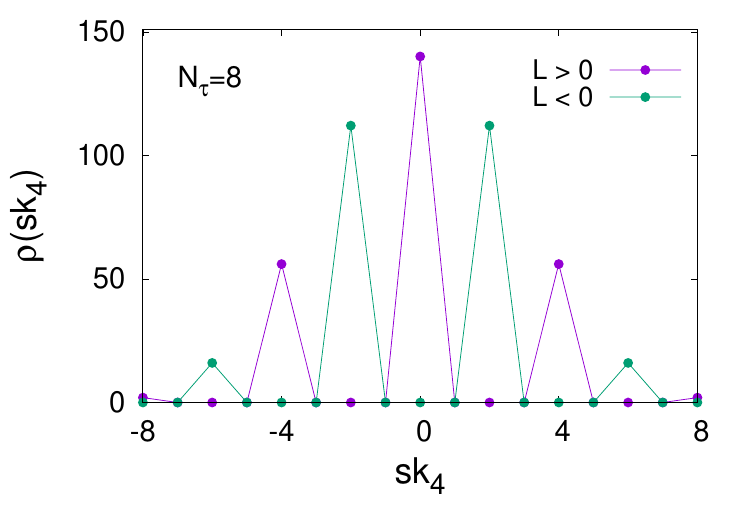}
    \caption{$\rho(sk_4)$ for $\kappa=0$ in 0+1}
	  \label{fig23}
  \end{minipage}

\centering
    \begin{minipage}[b]{0.45\textwidth}
    \includegraphics[width=\textwidth]{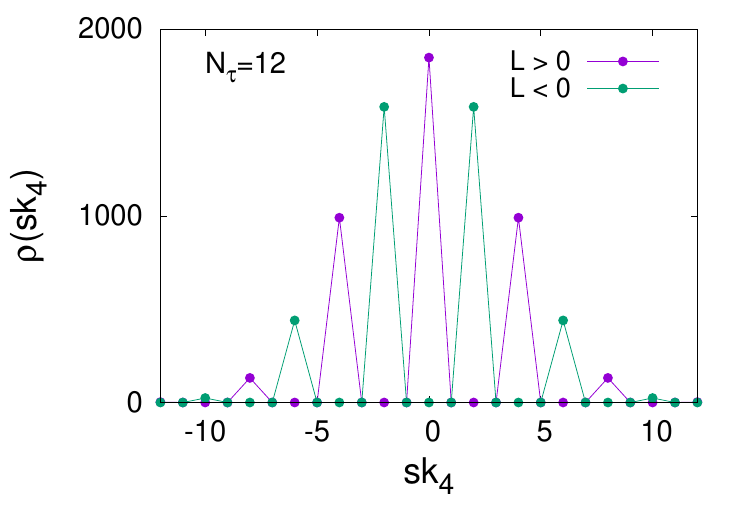}
    \caption{$\rho(sk_4)$ for $\kappa=0$ in 0+1}
    \label{fig24}
    \end{minipage}
  \hspace{0.3cm}
  \begin{minipage}[b]{0.45\textwidth}
    \includegraphics[width=\textwidth]{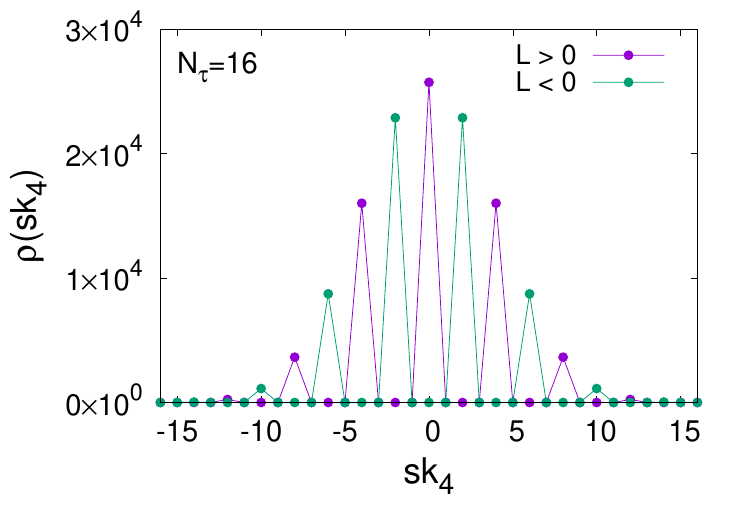}
    \caption{$\rho(sk_4)$ for $\kappa=0$ in 0+1}
   \label{fig25}
  \end{minipage}
 \end{figure}
 
 As noted previously the realisation of the $Z_2$ symmetry (vanishingly small explicit breaking) must come from the $Z_2$ symmetry of the entropy or the $DoS$.
  For $L=1$ the sequence of allowed  value of $sk_4$ is $\{N_\tau, N_\tau-4,......\ge -N_\tau\}$. On the other hand for $L=-1$
 the corresponding sequence is $\{N_\tau-2,N_\tau-6,........\ge 2-N_\tau\}$. The $DoS$ or $\rho(sk_4)$ for $N_\tau=4,8,12$
 and $16$ are shown in figures.\ref{fig22}-\ref{fig25}. For small $N_\tau$ there are clear difference for $L=\pm 1$.
The difference persists for the largest as well as smallest values of $sk_4$. For large $N_\tau$, $\rho(sk_4)$'s for both 
 $L=\pm 1$ are well described by a gaussian centred at $sk_4=0$, with $\sqrt{N_\tau}$ as standard deviation. The logarithm of the 
 peak hight is given by $~\simeq {\rm log}N_\tau!-2{\rm log}(N_\tau/2)! + {\rm log}2$ for $N_\tau$ even. For $N_\tau=2n+1$ the same can be approximated by $ {\rm log}N_\tau!-{\rm log}(n^2+n)+{\rm log2}$. The thermodynamics in the $N_\tau \to \infty$ limit will be dominated 
 by peak height and distribution of $\rho(sk_4)$ around the peak, which is $Z_2$ symmetric, for all finite $\kappa$. Interestingly this situation is similar to one dimensional Ising chain where entropy dominates for any non-zero finite temperature. 
 
\begin{figure}[!tbp]
\centering
    \begin{minipage}[b]{0.45\textwidth}
    \includegraphics[width=\textwidth]{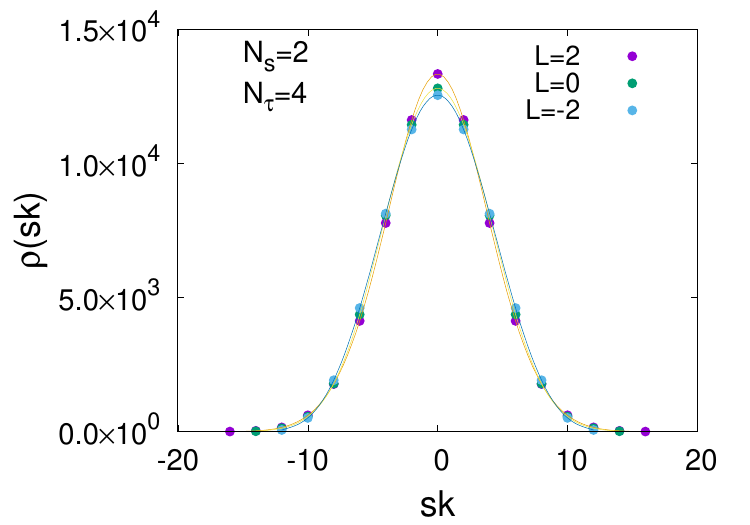}
    \caption{$\rho(sk)$ for $\kappa=0$ in 0+1}
   \label{fig20}	    
  \end{minipage}
  \hspace{0.3cm}
  \begin{minipage}[b]{0.45\textwidth}
    \includegraphics[width=\textwidth]{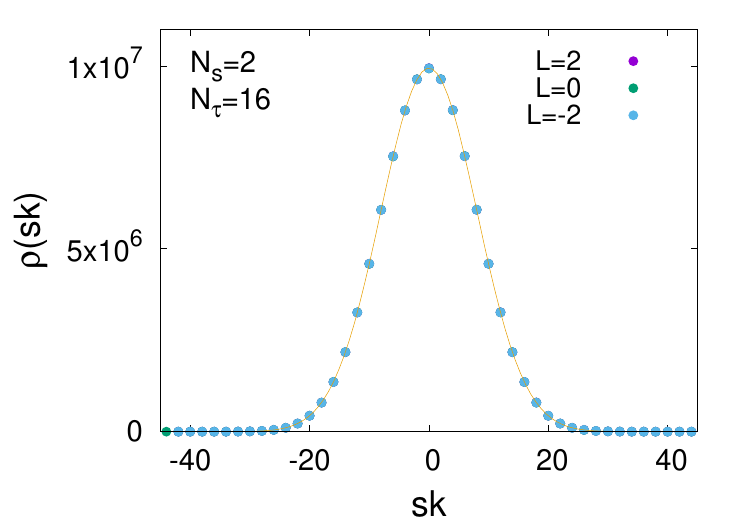}
    \caption{$\rho(sk)$ for $\kappa=0$ in 0+1}
	  \label{fig21}
  \end{minipage}
  \end{figure}

 In order to take into account the effect of nearest neighbour coupling along the spatial direction we consider $1+1$ dimensional
 model with $N_s=2$ and vary $N_\tau$. In this case the Polyakov loop can take value $L=0,\pm 2$. The exact calculation of 
 $\rho(sk)$ get increasingly difficult with $N_\tau$. One can however
 consider generating configurations randomly by giving equal probability for each allowed value of a given variable. The results for
 the distribution of the total action for 
 $N_\tau=4$ and $N_\tau=16$ are shown in Figs.\ref{fig20}-\ref{fig21}. As one can see that for higher $N_\tau$, $\rho(sk)$ around the peak $sk=0$ do not depend on $L$.
  
 To find out how well the $\rho(sk_4)$ describe the Monte Carlo simulations of the $4D$ partition function, the thermal average 
 of the distribution function $H(sk_4)$ of $sk_4$ has been computed. For each configuration $H(sk_4)$ is given by the number of spatial sites
 with a given value of $sk_4$. Note that the distribution of $sk_4$ takes into account the Boltzmann factor
  which shifts the peak of $\rho(sk_4)$ to the right. 
  \begin{figure}[!tbp]
\centering
    \begin{minipage}[b]{0.45\textwidth}
    \includegraphics[width=\textwidth]{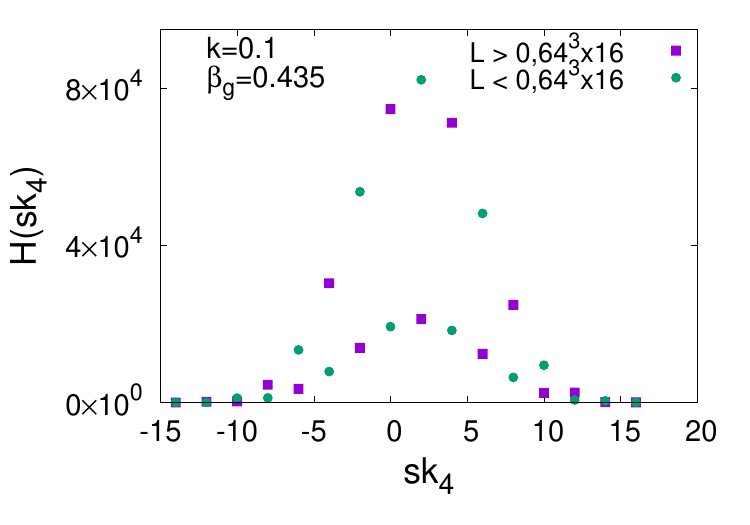}
    \caption{$H(sk_4)$ for $\kappa=0.1,~\beta_g =0.435$ for $3+1$ dimension}
   \label{fig22a}	    
  \end{minipage}
  \hspace{0.3cm}
  \begin{minipage}[b]{0.45\textwidth}
    \includegraphics[width=\textwidth]{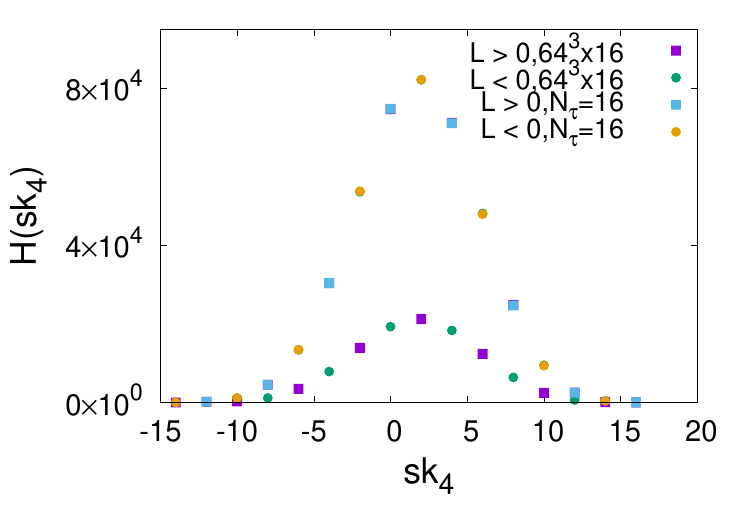}
    \caption{$H(sk_4)$ fitted with $0+1$ density of states with a Boltzmann factor}
	  \label{fig23b}
  \end{minipage} 
 \end{figure}
 The figure.~\ref{fig22a} shows the distribution $H(sk_4)$ for $N_\tau=16$ at $\kappa=0.1$ and $\beta_g =0.435$. For these values
 of $\kappa$ and $\beta_g$, the system is found to be in the deconfined and Higgs symmetric phase. The thermal average of
 the Polaykov loop for the two sectors are found to be $\left < L\right> = 0.5896\pm 0.002$ and $-0.5897\pm 0.00199$. Since the
 $\left<L\right> \ne 1$ there is a smaller but finite fraction of spatial site where the Polyakov loop takes opposite value. This results in the lower  envelope in $H(sk_4)$. The results clearly show that $H(sk_4)$ for both the Polyakov loop sectors can be approximately
described by single function in other words the presence of $Z_2$ symmetry. 
 
 In figure.~\ref{fig23b}, we try to fit the $3+1$ dimensional simulation result with $0+1$ dimensional $DoS$ by including an 
extra Boltzmann factor, i.e exp$(\kappa^\prime sk_4)$. The resulting fit agree very well with $H(sk_4)$. We expect that the $0+1$
results can describe the $3+1$ Monte Carlo simulations in most of the phase diagram except for critical points. Note here, $H(sk_4)$
values correspond to  $\kappa = 0.1$, however to fit $DoS$ one needs a $\kappa$ value which is higher. This is due to the fact that in $3+1$ dimensions $sk_4$ at a given spatial point interacts with $sk_4$ at the nearest neighbour sites. Considering
a mean-field approximation one can compute the free energy difference between $L=1$ and $L=-1$ at $\kappa=\kappa^\prime$  
for the $3+1$ dimensional system at $\kappa=0.1$, which turns out to be $10^{-10}$. 

\section{Conclusions}
\noindent 

In this paper the CD transition and $Z_2$ symmetry are studied in $Z_2$+Higgs theory in four dimensional space.
The results show that for large $N_\tau$ the $Z_2$ symmetry is realised in the Higgs symmetric phase within statistical errors. To understand the mechanism of emergence of the $Z_2$ symmetry a simplified one dimension model of $Z_2+$Higgs is 
considered by keeping only the temporal interaction terms at a given spatial site. The partition function and the corresponding free energy for each of the two Polyakov loop sectors is exactly calculated. It is shown that the free energy difference between the two Polyakov loop sectors vanishes in the large $N_\tau$ limit, which leads to $Z_2$ symmetry purely due to dominance of entropy. The $DoS$  for finite $N_\tau$ are calculated exactly where the asymmetry between the different Polyakov loop sectors rapidly decreases
with $N_\tau$. The effect of nearest neighbour interaction along the spatial directions in a simple model shows the persistence
of $Z_2$ symmetry in the $DoS$. Further it is shown that the $3+1$ Monte Carlo simulations can be reproduced using the $DoS$ of the one dimensional model. 

For a better understanding of the effects of $Z_2$ or $Z_N$ realisation on the confinement of static charges need to be studied
in $SU(N)$ gauge theories in view of the $Z_2+$Higgs results, which we plan to do in future. The realisation of $Z_N$ symmetry due to dominance of $DoS$, it's effect on the CD transition
and the $Z_N$ states in the deconfined phase will play an important role in the study of the early Universe.\\


\end{document}